\documentstyle[preprint,aps,epsfig,amsbsy,floats]{revtex}
 \tightenlines

\begin{document}
\newcommand{\beq}{\begin{equation}}
\newcommand{\eeq}{\end{equation}}
\newcommand{\beqa}{\begin{eqnarray}}
\newcommand{\eeqa}{\end{eqnarray}}
\newcommand{\sr}{\sqrt}
\newcommand{\fr}{\frac}
\newcommand{\mn}{\mu \nu}
\newcommand{\G}{\Gamma}

\draft \preprint{hep-th/0403210,~ INJE-TP-04-03}
\title{Second-order corrections to slow-roll inflation in the brane cosmology}
\author{ Kyong Hee Kim,  Hyung Won Lee and  Yun Soo Myung\footnote{E-mail address:
ysmyung@physics.inje.ac.kr}}
\address{
Relativity Research Center and School of Computer Aided Science\\
Inje University, Gimhae 621-749, Korea} \maketitle

\begin{abstract}
We calculate the power spectrum, spectral index, and running
spectral index for the RS-II brane inflation in the high-energy
regime  using the slow-roll expansion. There exist several
modifications. As an example, we take the power-law inflation by
choosing an inverse power-law potential. When comparing these with
those arisen in the standard inflation, we find that the power
spectrum is enhanced and the spectral index is suppressed, while
the running spectral index becomes zero as in the standard
inflation. However, since second-order corrections are rather
small, these could not play a role of distinguishing between
standard and brane inflations.
\end{abstract}

\thispagestyle{empty}
\setcounter{page}{0}
\newpage
\setcounter{page}{1}

There has been much interest in the phenomenon of localization of
gravity proposed by Randall and Sundrum (RS-II)~\cite{RS2}. They
assumed a single positive tension 3-brane and a negative bulk
cosmological constant in the five dimensional (5D) spacetime. They
obtained a 4D localized gravity by fine-tuning the tension of the
brane to the cosmological constant. Recently, several authors
studied cosmological implications of brane world scenario. The
brane cosmology contains some important deviations from the
Friedmann-Robertson-Walker (FRW) cosmology\cite{BDL1,BDL2}. The
Friedmann equation is modified at high-energy significantly.

On the other hand, it is generally accepted that curvature
perturbations produced during inflation are considered  to be
 the origin of  inhomogeneities necessary for
explaining  galaxy formation and other large-scale structure. The
first year results of WMAP put forward more constraints on
cosmological models and confirm the emerging standard model of
cosmology, a flat $\Lambda$-dominated universe seeded by
scale-invariant adiabatic gaussian fluctuations\cite{Wmap1}. In
other words, these results coincide with  predictions of the
standard inflation  with a single inflaton.  Also WMAP brings
about some new intriguing results: a running spectral index of
scalar metric perturbations and an anomalously low quadrupole of
the CMB power spectrum\cite{Wmap2}. If the brane inflation is
occurred, one expects that it gives us  different results in the
high-energy regime.

Recently Maartens et al.\cite{MWBH} have described the
inflationary perturbation in the RS-II brane cosmology using the
slow-roll approximation based on derivatives of the potential
driving inflation. Liddle and Taylor\cite{LT} have shown that in
the slow-roll approximation, the primordial perturbations alone
cannot be used to distinguish between an RS-II brane inflation and
standard inflation. More recently Ramirez and Liddle\cite{RL} have
studied the same issue using the slow-roll approximation based on
 derivatives of the Hubble parameters. They found that the
 first-order correction to the brane cosmology is of a similar size to that in
the standard cosmology. Also Tsujikawa and Liddle\cite{TL} have
investigated observational constraints on the RS-II brane
inflation from CMB anisotropies by introducing the large-field,
small-field, and hybrid models. Unfortunately, in the slow-roll
approximation\cite{SL}, there is no significant change in the
power spectrum  between the standard and brane cosmologies up to
first-order corrections\cite{Cal}. However,  in order to take into
account an effect of the brane inflation in the high-energy regime
correctly, we have to calculate the power spectrum up to second
order in the slow-roll parameters\cite{SG}. The slow-roll
approximation is not suitable for this purpose.

 In this brief
report, we will calculate the power spectrum, spectral index, and
running spectral index for the RS-II brane inflation using the
slow-roll expansion. We follow notations of Ref.\cite{RL} except
slow-roll parameters.

We start with a Friedmann equation in the RS-II brane cosmology by
adopting a flat FRW metric as a background spacetime on the brane
\begin{equation}
H^2=\fr{8\pi}{3M_4^2} \rho \Big(1+\fr{\rho}{2\lambda}\Big),
\end{equation}
where $H=\dot{a}/a,~M^2_4=1/G_4,~\lambda$ are the Hubble
parameter, 4D Planck mass, brane tension, respectively. Note the
relation between these and 5D Planck mass $M^3_5=1/G_5$:
$M_5^3=(4\pi \lambda/3)^{1/2}M_4$. Here we have set the 4D
cosmological constant to zero $(\Lambda_4=0$) and assumed that
inflation makes any holographic (dark)-radiation term
negligible\cite{holo}. In the limit of $\rho \ll \lambda$, we
recover the standard cosmology, while for $\rho \gg \lambda$, we
find the brane cosmology in the high-energy regime. Considering an
inflaton $\phi$ confined to the brane, one finds the equation
\begin{equation}
\ddot{\phi}+3H\dot{\phi}=-V^{\prime},
\end{equation}
where dot and prime denote   derivative with respect to time and
$\phi$, respectively. Its energy density and pressure are given by
$\rho=\dot{\phi}/2+V$ and $p=\dot{\phi}/2-V$. From now on we use
the following equations for the brane inflation:
\begin{equation}
H\simeq \fr{4\pi}{3M_5^3}V,~~3H\dot{\phi}\simeq
-V^{\prime},~~\dot{\phi}^2=-\fr{M_5^3}{4 \pi}\fr{\dot{H}}{H}.
\end{equation}
Here the first and second equations use the slow-roll
approximation. On the other hand the first and third equations use
the high-energy approximation and thus these are different from
those in the standard inflation ($H^2\simeq
\fr{8\pi}{3M_4^2}V,~\dot{\phi}^2=-\fr{M_4^2}{4 \pi}\dot{H}$). The
above enables us to introduce slow-roll parameters on the brane as
\begin{equation}
\epsilon_1 = -\frac{\dot H}{H^2},~~\delta_n \equiv
\frac{1}{H^n\dot{\phi}}\frac{d^{n+1}\phi}{dt^{n+1}}
\end{equation}
which satisfy the slow-roll condition:
$\epsilon_1<\xi,~|\delta_n|<\xi^n$ for some small perturbation
parameter $\xi$ defined on the brane.
 Here
 the subscript denotes the order in the slow-roll
expansion. When comparing these with those in Ref.\cite{RL}, one
finds relations that $\epsilon_1 \to \epsilon_H,~\delta_1 \to
-\eta_H$.

We are now in a position to calculate the perturbations using the
Mukhanov's formalism. Introducing a new variable $u=a\delta \phi$
where $ \delta \phi$ is a perturbed inflaton, its Fourier modes
$u_k$ in the linear perturbation theory satisfies the
Mukhanov-type equation\footnote{By comparison, we list standard
and brane potential-like terms: $\frac{1}{z}\frac{d^2z}{d\tau^2}=
2a^2H^2\Big( 1+\epsilon_1+\frac{3}{2}\delta_1+\epsilon_1^2
+2\epsilon_1\delta_1 +\frac{1}{2}\delta_2\Big)$ for standard
inflation and $\frac{1}{z}\frac{d^2z}{d\tau^2}= 2a^2H^2\Big(
1+\epsilon_1+\frac{3}{2}\delta_1+\frac{1}{2}\epsilon_1^2
+2\epsilon_1\delta_1 +\frac{1}{2}\delta_2\Big)$ for brane
inflation. Only one change in coefficient of $\epsilon_1^2$ occurs
: $1 \to \frac{1}{2}$.} \cite{muk}:
\begin{equation}
\label{eqsn} \frac{d^2u_k}{d\tau^2} +
\left(k^2-\frac{1}{z}\frac{d^2z}{d\tau^2}\right)u_k =
0,\end{equation} where $\tau$ is the conformal time defined by
$d\tau=dt/a$, and $z=a\dot{\phi}/H$ encodes all information about
the slow-roll inflation. Although this equation has been derived
in the standard cosmology, we still use it for a perturbation
study of the brane inflation\cite{RL}. Here $u_k$ depends on the
norm of ${\bf k}$ only because we work in an isotropic background.
In general its asymptotic solutions are obtained as
\begin{equation}\label{bc}
u_k \longrightarrow \left\{
\begin{array}{l l l}
\frac{1}{\sqrt{2k}}e^{-ik\tau} & \mbox{as} & -k\tau \rightarrow \infty \\
A_k z & \mbox{as} & -k\tau \rightarrow 0.
\end{array} \right.
\end{equation}
 The first solution corresponds to the flat space vacuum on
scale much smaller than the Hubble distance (sub-horizon scale),
and the second is a growing mode on scale much larger than the
Hubble distance (super-horizon scale). Using  a definition of the
power spectrum for the curvature perturbation defined by $R_{c{\bf
k}}=-u_{\bf k}/z$ with $u_{\bf k}(\tau)=a({\bf
k})u_k(\tau)+a^{\dagger}(-{\bf k})u^*_k(\tau)$:
$P_{R_c}(k)\delta^{(3)}({\bf k}-{\bf q})=\fr{k^3}{2\pi^2}<R_{c{\bf
k}}(\tau)R^{\dagger}_{c{\bf q}}(\tau)>$, one finds
\begin{equation}\label{gps}
P_{R_c}(k) = \left(\frac{k^3}{2\pi^2}\right)
\lim_{-k\eta\rightarrow0}\left|\frac{u_k}{z}\right|^2 =
\frac{k^3}{2\pi^2}|A_k|^2.
\end{equation}
Our task is to find $A_k$ by solving the Mukahnov-type equation
(\ref{eqsn}). In general it is hard to solve it. However, we can
solve it using either the slow-roll approximation \cite{SL} or the
slow-roll expansion\cite{SG}.
 In the slow-roll approximation  we take  $\epsilon_1$ and $\delta_1$ to be constant.
Thus the slow-roll approximation could not be considered as a
general approach beyond the first-order correction to calculate
the power spectrum\cite{KLM,KLLM}. In order to calculate the power
spectrum up to second-order correctly, one should use the
slow-roll expansion based on Green's function technique. The key
step is to use a variable nature of slow-roll parameters: \beq
\label{slow-di}\dot \epsilon_1
=H(\epsilon_1^2+2\epsilon_1\delta_1),~~\dot{\delta}_1=
H(\epsilon_1\delta_1-\delta^2_1+\delta_2),\cdots \eeq which means
that  derivative of slow-roll parameters with respect to time
increases their  order by one in the slow-roll expansion. After a
lengthly calculation following ref.\cite{SG}, we find the brane
power spectrum
\begin{eqnarray}
 \label{2ndps}
P^{b}_{R_c}(k) & = &  \fr{H^4}{(2\pi)^2\dot{\phi}^2} \left\{ 1
-2\epsilon_1 + 2\alpha(2\epsilon_1+\delta_1)+
\left(6\alpha^2-2\alpha-21+\frac{13\pi^2}{6}\right)\epsilon_1^2 \right.\\
&& \left. +
\left(3\alpha^2+2\alpha-22+\frac{29\pi^2}{12}\right)\epsilon_1\delta_1
+ \left(3\alpha^2-4+\frac{5\pi^2}{12}\right)\delta_1^2 +
\left(-\alpha^2+\frac{\pi^2}{12}\right)\delta_2 \right\} \nonumber
\end{eqnarray}
and the right hand side should be evaluated at horizon crossing of
$k=aH$. Here $\alpha$ is defined by $\alpha=2-\ln2-\gamma \simeq
0.7296$ where $\gamma$ is the Euler-Mascheroni constant, $\gamma
\simeq 0.5772.$  Comparing with that from the standard
inflation\cite{SG}, we find a slight change in coefficient of
$\epsilon^2_1$: $2.15856 \to 2.11902$. We identify this with an
effect of the brane inflation in high-energy regime. Using
 $\fr{d\epsilon_1}{d\ln k}=
\fr{\epsilon_1^2+2\epsilon_1\delta_1}{1-\epsilon_1},~\fr{d\delta_1}{d\ln
k}=\fr{\epsilon_1\delta_1-\delta_1^2+\delta_2}{1-\epsilon_1},~\fr{d\delta_2}{d\ln
k}=\fr{2\epsilon_1\delta_2-\delta_1\delta_2+\delta_3}{1-\epsilon_1}$,
and $\fr{d\delta_3}{d\ln
k}=\fr{3\epsilon_1\delta_3-\delta_1\delta_3+\delta_4}{1-\epsilon_1}$,
the brane spectral index defined by
\begin{equation}
n_{s}^b(k) = 1 + \frac{d \ln P^{b}_{R_c}}{d \ln k}
\end{equation}
can be easily calculated up third order

\begin{eqnarray}
\label{11} n^b_{s}(k) =& & 1 - 4\epsilon_1 - 2\delta_1
+(4\alpha-6)\epsilon_1^2
              + (10\alpha -6)\epsilon_1\delta_1 -2\alpha\delta_1^2+2\alpha\delta_2
              \\ \nonumber
          & & + \left(
                    -4\alpha^2+16\alpha-52+{13\pi^2\over 3}
                \right)\epsilon_1^3
                +\left(
                    -18\alpha^2+46\alpha-142+{72\pi^2\over 2}
                \right)\epsilon_1^2\delta_1
                \\ \nonumber
          & & + \left(
                    -3\alpha^2+4\alpha-30+{13\pi^2\over 4}
                \right)\epsilon_1\delta_1^2
                +\left(
                    -7\alpha^2+8\alpha-22+{31\pi^2\over 12}
                \right)\epsilon_1\delta_2
                \\ \nonumber
          & & + \left(
                    -2\alpha^2+8-{5\pi^2\over 6}
                \right)\delta_1^3
                +\left(
                    3\alpha^2-8+{3\pi^2\over 4}
                \right)\delta_1\delta_2
                +\left(
                    -\alpha^2+{\pi^2\over 12}
                \right)\delta_3.
\end{eqnarray}
Here we find three changes in $\epsilon_1^2,~\epsilon^3_1$ and
$\epsilon^2_1 \delta_1$ : $-2.1629 \to -3.08145$,~ $4.48387 \to
0.312997$ and $18.9457 \to 15.2203$, respectively.
 Finally the brane running
spectral index up to fourth order is given by

\begin{eqnarray}
\label{12} \frac{d}{d\ln k} n^b_{s} & = &
-4\epsilon^{2}_{1}-10\epsilon_1\delta_1+2\delta^{2}_{1}-2\delta_2+(8\alpha
-16)\epsilon^{3}_{1}+(36\alpha -46)\epsilon^{2}_{1}\delta_1
\\ \nonumber
&+&(6\alpha-4)\epsilon_1\delta^{2}_{1}+(14\alpha
-8)\epsilon_1\delta_2+4\alpha\delta^{3}_{1}-6\alpha\delta_1\delta_2+2\alpha\delta_3 \\
\nonumber
         &+&\left(-12\alpha^2+56\alpha-172+13\pi^2\right)\epsilon_1^4
            +\left(-78\alpha^2+270\alpha-784+{133\pi^2\over
             2}\right)\epsilon_1^3\delta_1
               \\ \nonumber
         &+&\left(-63\alpha^2+156\alpha-520+{201\pi^2\over 4}
             \right)\epsilon_1^2\delta_1^2
            +\left(-39\alpha^2+84\alpha-216+{85\pi^2\over
             4}\right)\epsilon_1^2\delta_2
               \\ \nonumber
         &+&\left(-6\alpha^2+4\alpha+24-{5\pi^2\over
             2}\right)\epsilon_1\delta_1^3
            +\left(-4\alpha^2+10\alpha-106+{34\pi^2\over
             3}\right)\epsilon_1\delta_1\delta_2
               \\ \nonumber
         &+&\left(-10\alpha^2+10\alpha+22-{17\pi^2\over
             6}\right)\epsilon_1\delta_3
            +\left(6\alpha^2-24+{5\pi^2\over 2}\right)\delta_1^4
               \\ \nonumber
         &+&\left(-12\alpha^2+40-4\pi^2\right)\delta_1^2\delta_2
            +\left(4\alpha^2-8+{2\pi^2\over 3}\right)\delta_1\delta_3
               \\ \nonumber
         &+&\left(3\alpha^2-8{3\pi^2\over 4}\right)\delta_2^2
            +\left(-\alpha^2+{\pi^2\over 12}\right)\delta_4.
            \nonumber
\end{eqnarray}
Here we have several changes: $-8 \to -4(\epsilon_1^2)$, $-16.6516
\to -10.1629(\epsilon_1^3)$, $-16.2218 \to -19.7331(\epsilon_1^2
\delta_1)$, $12.0516 \to 9.22391(\epsilon_1^4)$, $108.669 \to
27.8058(\epsilon_1^3 \delta_1)$, $70.8055 \to 56.2317(\epsilon_1^2
\delta_1^2)$, $48.5885\to 34.2562(\epsilon_1^2 \delta_2)$.

  As a concrete
example, we choose the power-law inflation like $a(t)\sim t^p$
whose potential is given by\cite{RL,Cal} \beq
V^b(\phi)=\fr{M_5^6(6p-1)}{8\pi^2}\fr{1}{\phi^2},~~p>1.\eeq For
$M_4=\sqrt{8\pi}$, this gives us $\frac{V^b}{\lambda}=
\frac{4(6p-1)}{3}\frac{1}{\phi^2}$. Instead of an exponential
potential $V^s=V_0 \exp(-\sqrt{2/p} \phi)$ for the standard
inflation, here we choose an inverse power-law potential. We note
that for $V_0=4(6p-1)/3$, two potentials of $V^b/\lambda$ and
$V^s$ take similar shapes in the slow-roll period. Then brane
slow-roll parameters\footnote{On the other hand, the potential
slow-roll parameters in the high-energy limit are given by
$\epsilon^V_1=\frac{3M_5^6}{16\pi^2}\frac{V'^2}{V^3}=\frac{6}{6p-1},
~\delta^V_1=-\frac{3M^6_5}{16\pi^2}\Big[\frac{V"}{V^2}-\frac{V'^2}{V^3}\Big]
=-\frac{3}{6p-1}$. Similarly, we obtain
$\delta_2^V=\frac{27}{(6p-1)^2},~\delta_3^V=-\frac{405}{(6p-1)^3},
~\delta_4^V=\frac{8505}{(6p-1)^4}$. Hence, for large $p>1$, we
have approximate relations: $\epsilon_1 \simeq
\epsilon_1^V,~\delta_n \simeq \delta_n^V$.} are determined by \beq
\label{pislp} \epsilon_1=\fr{1}{p},~\delta_1=-\fr{1}{2p},~
\delta_2=\fr{3}{4p^2},~\delta_3=-\fr{15}{8p^3}
,~\delta_4=\fr{105}{16p^4}\eeq which are obtained from relations
in Eq.(\ref{slow-di}) after setting $\epsilon_1=1/p$\cite{RL}.

For a reference, we list up the exact form of slow-roll parameters
in the standard power-law inflation \beq \label{spislp}
\epsilon_1=\fr{1}{p},~\delta_1=-\fr{1}{p},~
\delta_2=2\delta_1^2=\fr{2}{p^2},~\delta_3=6\delta_1^3=-\fr{6}{p^3}
,~\delta_4=24\delta_1^4=\fr{24}{p^4}.\eeq The brane power spectrum
takes the form \beqa
 \label{2ndpssb}
P^{PI,b}_{R_c}(k)& = &\frac{H^4}{(2\pi)^2\dot\phi^2} \left\{ 1
+\Big(3\alpha-2\Big)\fr{1}{p}+\Big(\fr{9}{2}\alpha^2
-3\alpha-11-\fr{9\pi^2}{8} \Big)\fr{1}{p^2}
        \right\} \\ \nonumber
       & =&\frac{H^4}{(2\pi)^2\dot\phi^2} \left\{ 1
+\fr{0.1888}{p}+\fr{0.309928}{p^2}
        \right\}. \eeqa
On the other hand, the power spectrum in standard power-law
inflation  is given by

\beqa\label{2ndpss} P^{PI,s}_{R_c}(k)& =&
\frac{H^4}{(2\pi)^2\dot\phi^2} \left\{ 1
+2\Big(\alpha-1\Big)\fr{1}{p}+\Big(2\alpha^2
-2\alpha-5-\fr{\pi^2}{2} \Big)\fr{1}{p^2}
        \right\} \\ \nonumber
        &=&\frac{H^4}{(2\pi)^2\dot\phi^2} \left\{ 1
-\fr{0.540726}{p}-\fr{0.459731}{p^2}
        \right\}. \nonumber \eeqa
The brane spectral index can be easily calculated up to third
order
 \beq  n^{PI,b}_{s}(k)=1-\fr{3}{p}-\fr{3}{p^2}-\fr{3}{p^3}.
\eeq whereas the spectral index in standard power-law inflation is
given by \beq n^{PI,s}_{s}(k)=1-\fr{2}{p}-\fr{2}{p^2}-\fr{2}{p^3}.
\eeq Finally the brane running spectral index is found to be zero
up to $1/p^4$ as the running spectral index in the standard
power-law inflation did, \beq \frac{d n^{PI,b}_s}{d\ln k}=\frac{d
n^{PI,s}_s}{d\ln k}=0. \eeq

We calculate the power spectrum, spectral index, and running
spectral index for the RS-II brane inflation in the high-energy
regime  using the slow-roll expansion. When using the slow-roll
approximation, it appears that there is no change in power
spectrum up to first-order corrections  between  brane and
standard inflations. However, as are shown in Eqs.(\ref{2ndps}),
(\ref{11}), and (\ref{12}), there are several modifications
compared with those of the standard inflation. As an example, we
take the power-law inflation by choosing an inverse power-law
potential. When comparing these with those arisen in the standard
inflation, we find that the power spectrum is enhanced and the
spectral index is suppressed, while the running spectral index
becomes zero as in the standard inflation. Explicitly, choosing
$p=101$ \footnote{In Ref.\cite{RL}, the authors choose $p=101$ to
take $N=50$ $e$-foldings before the end of inflation. However,
they use a potential of $V=\phi^{\alpha},~\alpha=2,4,6$ for brane
and standard inflations to obtain fractional corrections to the
power spectrum for standard and brane cosmology. As was previously
pointed out, the power-law inflation can be achieved only  when
using $V^b \propto \phi^{-2} (\exp(-\sqrt{2/p}\phi))$ for brane
(standard) inflations. Hence our comparison test is suitable for
the power-law inflation.} leads to
$\frac{(2\pi)^2\dot\phi^2}{H^4}P^{PI,b}=1,~1.000187,~1.00019$ for
zeroth, first, second-order corrections, respectively, whereas
$\frac{(2\pi)^2\dot\phi^2}{H^4}P^{PI,s}=1,~0.994646,~0.99461$.
Also we find the spectral index that
$n_s^{PI,b}=0.970297,0.970003,~0.97$ for zeroth, first,
second-order corrections, respectively, while
$n_s^{PI,s}=0.980198,~0.980002,~0.98$. According to WMAP data,
power spectrum normalization at $k_0=0.05 {\rm Mpc}^{-1}$ is given
by $A=0.833^{+0.086}_{-0.083}$ and scalar spectral index is
$n_s=0.93^{+0.03}_{-0.03}$ at $k_0=0.05 {\rm Mpc}^{-1}$.
Apparently we obtain blue (red) power spectrum corrections for
brane (standard) inflation. This is not a crucial result because
we measure only normalization factor of power spectrum from WMAP.
Also we have red spectral index for both cases. Since recent
observations including WMAP have restricted viable inflation
models to regions close to the slow-roll limit, our second-order
corrections to the slow-roll brane inflation in the high-energy
limit are rather small. Hence we confirm that in the slow-roll
approximation, observations of the primordial perturbation spectra
cannot distinguish between RS-II scenario and standard
inflation\cite{LT}.

We conclude that our second-order corrections which are even
slightly different from those of the standard inflation could not
play a role in distinguishing between  RS-II brane inflation and
standard inflation. The two main reasons are as follows. First,
the two models are based on the same perturbation equation given
by Eq.(\ref{eqsn}) with slightly different potential-like terms:
$\frac{1}{z}\frac{d^2z}{d\tau^2}$ (see footnote 1). Although there
is no rigorous derivation of this Mukhanov-type equation in the
brane inflation, we adopt it for simplicity. Second, the brane
cosmology in the high-energy limit belongs to the patch
cosmology\cite{Cal}: $H^2 \propto \rho^q$ with $q=1$ for standard
cosmology (SD), $q=2/3$ for high-energy Gauss-Bonnet (GB) and
$q=2$ for high-energy RS-II. The patch cosmology with an inflaton
gives us similar results in the slow-roll expansion except a
relation of $\dot \epsilon_1
=2H(\epsilon_1^2/q+\epsilon_1\delta_1)$. Also we have the same
power-law inflation even for different potentials: $\phi^6$ for
GB, $\exp(-\sqrt{2/p}\phi)$ for SD, and $\phi^{-2}$ for RS-II.

\subsection*{Acknowledgements}
We thank Hungsoo Kim, Gil Sang Lee  and G. Calcagni for helpful
discussions. Y.S. was supported in part by KOSEF, Project No.
R02-2002-000-00028-0. H.W. was in part supported by KOSEF,
Astrophysical Research Center for the Structure and Evolution of
the Cosmos.

\end{document}